\documentstyle[tighten,floats,aps,epsf]{revtex}

\begin{document}
\title
{Dipole resonances in light neutron-rich nuclei studied 
with time-dependent calculations of antisymmetrized molecular dynamics}

\author{Y. Kanada-En'yo and M. Kimura}

\address{Yukawa Institute for Theoretical Physics, Kyoto University,\\
Kyoto 606-8502, Japan}

\maketitle
\begin{abstract}
In order to study isovector dipole response of neutron-rich nuclei,
we have applied a time-dependent method of antisymmetrized 
molecular dynamics.
The dipole resonances in Be, B and C isotopes have been investigated.
In $^{10}$Be, $^{15}$B, $^{16}$C, 
collective modes of the 
vibration between a core and valence neutrons 
cause soft resonances at the excitation energy $E_x=10-15$ MeV
below the giant dipole resonance(GDR).
In $^{16}$C, we found that a remarkable peak at $E_x=14$ MeV corresponds to 
coherent motion of four valence neutrons against a $^{12}$C core, 
while the GDR arises from the core vibration in the $E_x >20$ MeV region.
In $^{17}$B and $^{18}$C, the dipole strengths in the low
energy region decline compared with those in $^{15}$B and $^{16}$C.  
We also discuss the energy weighted sum rule for the $E1$ transitions.
\end{abstract}

\noindent

\section{Introduction}
In neutron-rich nuclei, there often appear 
exotic phenomena which are very different from those in stable nuclei 
due to the excess neutrons. Some properties of neutron-rich nuclei
are concerned with differences between 
proton and neutron densities.
Neutron halo and skin structures are typical examples.
Another related subject is deformations of proton and neutron densities. 
For example, the difference of the deformations between 
proton and neutron densities were theoretically suggested 
in some Be, B and C isotopes \cite{ENYObc,ENYO-c10,ENYO-c16}. 
These phenomena imply that the structure of 
the nuclei far from the $\beta$-stability line often contradicts to 
the usual understanding for 
stable nuclei where the proton and neutron densities are
consistent with each other in a nucleus. They may lead also to exotic 
phenomena in excitations and reactions.
One of the current issues is the dipole excitations 
in neutron-rich nuclei
\cite{Honma90,Hoshino91,Sagawa92,Tohyama95,Hamamoto96a,Hamamoto96b,Catara,Hamamoto98,Myo98,Sagawa-o,Suzuki00,Sagawa-be,Colo,Sagawa-c,Nakatsukasa,Nakamura,Leistenschneider,Palit}. 
Since it is quite natural to expect the
low-energy isovector-dipole excitations due to the difference between 
the proton and neutron densities, interests are attracted 
to the soft resonances below the giant dipole 
resonance(GDR), their collectivity and 
the contributions of the excess neutrons.
In fact, the features of dipole transitions in 
neutron-rich O isotopes have been studied 
experimentally\cite{Leistenschneider} and theoretically 
\cite{Hamamoto98,Sagawa-o,Colo}. 
They have been found to be different from those in 
the stable nucleus, $^{16}$O, especially in the low energy region below the 
GDR. The dipole excitations 
have been studied also in C isotopes 
by Suzuki et al. with shell model calculations\cite{Sagawa-c}, 
where they suggested that 
coherent neutron transitions enhance the strengths 
at the excitation energy
$E_x = 10-15$ MeV.  

Our present interest is in the isovector dipole excitations in the 
light neutron-rich nuclei and in the effect of the ground state properties like
the deformations on the $E1$ strengths.
A method of antisymmetrized molecular dynamics(AMD)\cite{ENYObc}
is one of powerful approaches for nuclear structure study.
The method is superior especially in 
the description of cluster aspect, which is important in light unstable nuclei
as well as in light stable nuclei\cite{ENYOsup}. 
In the systematic studies of Be, B, and C isotopes performed with
the AMD method,
a variety of structure such as the neutron skin and the deformed states 
have been suggested in those nuclei, and some of them have 
been discussed in relation to cluster 
aspect\cite{ENYObc,ENYOsup}. The experimental data for 
various properties of the neutron-rich Be, B and C isotopes have been 
successfully reproduced by 
the AMD calculations.
We should stress that the AMD calculations well agree to the 
experimental data of quadrupole moments and 
$E2$ strengths in neutron-rich B and C, 
which can not be reproduced by the shell model calculations 
without using system-dependent effective charges. 
For the study of dipole excitations in the AMD framework,
we apply a time-dependent method and calculate the
dipole strengths in a similar way to the the time-dependent 
Hartree-Fock(TDHF). The point is that we are able to 
study dipole resonances with the framework which can describe 
cluster aspect.
One of the advantages of the time-dependent
AMD is that we can link the excitations with such collective modes as 
core vibration, core-neutron motion and inter-cluster motion
which should be important 
to understand the role of the excess neutrons in the dipole resonances.

The time-dependent method of AMD have been proposed 
and applied to heavy-ion reactions by Ono et al. in 1992\cite{Ono92a,Ono92b}
earlier than the application of the AMD 
to the nuclear structure study.
For collective motion on the static solution, however,
the time-dependent AMD calculations
have not been performed yet. In this paper, we formulate a method
based of the time-dependent AMD for the study of 
$E1$ response in analogy to the TDHF. 
In order to see its validity, we first apply it to
$^{12}$C and $^{18}$O, 
and show comparison with the experimental data
and other theoretical calculations.
Then we apply this method to Be, B 
and C isotopes and discuss the properties of dipole strengths 
in the neutron-rich nuclei. We try to see how 
the dipole strength distribution is influenced by
such structure as deformations, neutron skin, 
the existence of core and clusters.

This paper is organized as follows. In the next section, we explain 
the formulation of the present method
for the $E1$ response, which is based on the time-dependent 
AMD. Adopted effective nuclear forces are described 
in \ref{sec:interaction}. In \ref{sec:results}, 
we show the results of $^{12}$C, $^{18}$O
and the comparison with the experimental data,
and also the dipole transitions in Be, B, C isotopes.
The discussion of
the $E1$ excitations in neutron-rich Be, B, and C isotopes 
are given in \ref{sec:discuss}.
Finally, in \ref{sec:summary} we give a summary.

\section{Formulation}
 \label{sec:formulation}
We explain the formulation of the time-dependent version of AMD 
for study of isovector dipole excitations.
By simulating time evolution of the
collective motion on the static solution with the time-dependent AMD, 
we can calculate the response of a nucleus to external dipole fields and
obtain the dipole strengths in the similar way to 
TDHF approaches.
 
The time-dependent method in the AMD framework 
is described in Refs.\cite{Ono92a,Ono92b}, where the method has been applied
to heavy-ion collisions. 
Concerning nuclear structure study with AMD methods, 
the static version of AMD and its extended versions are reviewed in Refs.
\cite{ENYOsup,AMDrev}.

\subsection{Wave function}
The wave function for a $A$-nucleon system($A$ is a mass number)
is given by a single Slater determinant of Gaussian wave packets as,
\begin{equation}
 \Phi = \frac{1}{\sqrt{A!}} {\cal{A}} \{
  \varphi_1,\varphi_2,...,\varphi_A \},
\end{equation}
where the $i$-th single-particle wave 
function is written as follows,
\begin{eqnarray}
 \varphi_i&=& \phi_{{\bf Z}_i}\chi_i\tau_i,\\
 \phi_{{\bf Z}_i}({\bf r}_j) &=& (\frac{2\nu}{\pi})^{3/4}
\exp\bigl\{-\nu({\bf r}_j-\frac{{\bf Z}_i}{\sqrt{\nu}})^2\bigr\},
\label{eq:spatial}\\
 \chi_i &=& (\frac{1}{2}+\xi_i)\chi_{\uparrow}
 + (\frac{1}{2}-\xi_i)\chi_{\downarrow}.
\end{eqnarray}
Here, the spatial part of the $i$-the single particle
wave function is given by a located Gaussian wave packet, whose center 
is represented by the complex parameter, 
${\bf Z}_i$.
The parameter $\xi_i$ indicates the orientation of the intrinsic spin, and
the iso-spin function is up(proton) or down(neutron).

In the present work, the orientation of the intrinsic spin is fixed to be 
up or down as $\xi_{i}=\{1/2,-1/2\}$ for simplicity.
In this AMD wave function, all the centers of Gaussian wave 
packets for $A$ nucleons are
independent variational parameters, and a set of 
parameters ${\bf Z}\equiv
\{{\bf Z}_{1}$, ${\bf Z}_{2}$, $\cdots$, ${\bf Z}_{A}\}$ 
specifies the total wave function $\Phi({\bf Z})$ of the state.
This is the simplest version of AMD wave function, and
the parity and angular momentum projections are not performed in the present
work.

\subsection{Equation of motion}
In the time-dependent version of AMD, 
components of ${\bf Z}$ are considered to be time-dependent parameters 
as explained in \cite{Ono92b}. 
The time evolution of the system is 
determined by the time-dependent variational principle,
\begin{equation}
\delta \int^{t_2}_{t_1} dt 
\frac{
\langle\Phi({\bf Z})|i\hbar\frac{d}{dt}-H|
\Phi({\bf Z})\rangle}{\langle\Phi({\bf Z})|\Phi({\bf Z})\rangle}=0.
\end{equation}
This leads to the equation of motion with respect to ${\bf Z}$,
\begin{equation}\label{eq:eqmotion}
i\hbar\sum_{j,\tau} C_{i\sigma,j\tau} \dot{Z}_{j\tau}=
\frac{\partial {\cal H}}{\partial Z^*_{i\sigma}} {\quad \rm and \quad c.c.}, 
\end{equation}
where $\sigma, \tau=x,y,z$, and ${\cal H}$ is the expectation value of the 
Hamiltonian $H$,
\begin{equation}\label{eq:hamil}
{\cal H}({\bf Z},{\bf Z}^*)=
\frac{
\langle\Phi({\bf Z})|H|
\Phi({\bf Z})\rangle}{\langle\Phi({\bf Z})|\Phi({\bf Z})\rangle}.
\end{equation}
\begin{equation}\label{eq:cmat}
C_{i\sigma,j\tau}\equiv 
\frac{\partial^2}{\partial Z^*_{i\sigma}\partial Z_{j\tau}}
\ln \langle\Phi({\bf Z})|\Phi({\bf Z})\rangle
\end{equation}
is a positive definite Hermitian matrix.
These equations, (\ref{eq:eqmotion}), (\ref{eq:hamil}) and (\ref{eq:cmat})
are derived in general from the time-dependent variational principle 
for a given wave function parametrized by complex variational parameters.
In case of the AMD framework, the time evolution of a system is 
described by the motion of the centers of Gaussian wave packets.

Although stochastic collision process has been introduced in the
study of heavy-ion collisions\cite{Ono92b}, 
we do not put it in the present framework.

\subsection{Response to dipole fields}
In order to calculate the response to external fields, 
we first solve the static problem to obtain the optimum solution 
$\Psi^0$ for the ground state. We perform the energy variation 
of AMD wave function with respect 
to the variational parameter ${\bf Z}$ 
by using the frictional cooling method(a imaginary-time method)
\cite{ENYOsup,Ono92a,Ono92b}. We obtain the optimum parameter
${\bf Z}^0$, which gives the energy minimum state $\Psi^0=\Phi({\bf Z}^0)$ 
in the AMD model space.
Then, we boost the $\Psi^0$ instantaneously at t=0
by imposing an external perturbative field, 
$V_{\rm ext}({\bf r},t) = \epsilon F({\bf r})\delta(t)$, where $\epsilon$
is an arbitrary small number. 
This results in an initial
state of the time-dependent calculation as follows:
\begin{equation}
\Psi(t=0+)=e^{-i\epsilon F}\Psi^0=e^{-i\epsilon F}\Phi({\bf Z}^0) 
\end{equation}
In the calculation of $E1$ resonances, the external field
is chosen to be the dipole field as,
\begin{equation}
F({\bf r})=\epsilon {\cal M}(E1,\mu)=
\sum_i^A e^{\rm rec} r_iY_{1\mu}(\hat {\bf r_i}),
\end{equation}
where $e^{\rm rec}$ is the $E1$ recoil charge, $Ne/A$ for protons and 
$-Ze/A$ for neutrons. Then the initial state $\Psi(t=0+)$ is written with 
a single AMD wave function $\Phi({\bf Z}(t=0+))$ 
by simply transforming the parameters ${\bf Z}^0=\{{\bf Z}^0_1,
{\bf Z}^0_2,\cdots,{\bf Z}^0_A\}$ as
follows:
\begin{equation}
{\bf Z}_i(t=0+)={\bf Z}_i^0-\frac{\epsilon e^{\rm rec} 
\mbox{\bf e}_\mu}{2\sqrt{\nu}}i,
\end{equation}
where {\bf e}$_\mu$ is the unit vector. 
Although an extra normalization factor of the
wave function arises from this transformation, it gives no effect
on the physical quantities because the AMD framework is always 
based on the normalized wave functions.

By using the equation of motion (Eq.(\ref{eq:eqmotion})),
we can calculate the time evolution of the system, 
$\Psi(t)=\Phi({\bf Z}(t))$, from the initial
state $\Psi(t=0+)=\Phi({\bf Z}(t=0+))$ following the time-dependent AMD.
The transition strength is obtained by 
Fourier transform of the expectation value of ${\cal M}(E1,\mu)$ as follows,
\begin{equation}\label{eq:BE1}
\frac{d B(\omega;E1,\mu)}{d\omega}\equiv \sum_n 
|\langle n|{\cal M}(E1,\mu)|0 \rangle|^2 \delta(\omega-\omega_n) = 
-\frac{1}{\pi\epsilon} {\rm Im}\int^\infty_0 dt \langle \Psi(t)| {\cal M}(E1,\mu) 
|\Psi(t) \rangle e^{i\omega t},
\end{equation}
where $|0\rangle$ is the ground state and 
$|n\rangle$ is the excited state with the excitation energy $\hbar\omega_n$.
In the deformed nuclei, Eq. (\ref{eq:BE1}) gives the 
$E1$ transition strengths in the intrinsic state 
because the total angular momentum projection is not performed.
Assuming the strong coupling scheme, we calculate the $B(E1)$ 
in the laboratory frame by sum of the 
intrinsic $E1$ strengths as follows:
\begin{equation}
\frac{d B(\omega;E1)}{d\omega}= \sum_{K=0,\pm 1}
\frac{d B(\omega;E1,K)}{d\omega}.
\end{equation}
In the practical calculation, we impose the $E1$ field with respect to 
each direction, $x,y,z$, independently, and sum up the strengths
instead of the sum of $K=0,\pm 1$. 
In the present framework, 
$d B(\omega;E1)/d\omega$ consists of 
discrete peaks in principle, because the present AMD is a bound state
approximation and continuum states are not taken into account.
We introduce a smoothing parameter $\Gamma$, add an 
imaginary part $i\Gamma/2$ to the real excitation energy $E_x$ as 
$\hbar \omega=E_x+i\Gamma/2$, and calculate the $B(E1)$ 
with Eq.(\ref{eq:BE1}) by performing the integral up to finite time.
This smoothing can be considered to simulate 
the escape and the spreading widths of the resonances.

The photonuclear cross section $\sigma(\omega)$ is related to the
transition strength $B(\omega;E1)$ as,
\begin{equation}
\sigma(\omega)=\frac{16\pi^3 }{9\hbar c} 
\hbar\omega\frac{d B(\omega;E1)}{d\omega}.
\end{equation} 

\section{Effective nuclear Interactions} 
\label{sec:interaction}

We use an effective nuclear interaction 
which consists of the central force, the
 spin-orbit force and the Coulomb force.
In the present work, we adopt MV1 force \cite{TOHSAKI} as the central force.
This force contains a zero-range three-body force 
in addition to the finite-range 
two-body interaction:
\begin{eqnarray}
V^{MV1}&=&\sum_{i<j} V^{(2)}+\sum_{i<j<k} V^{(3)},\\
V^{(2)}&=& (w+bP_\sigma-hP_\tau-m P_\sigma P_\tau)\left[
V_1\exp\left(-\frac{r_{ij}^2}{a_1^2} \right)
+V_2\exp\left(-\frac{r_{ij}^2}{a_2^2} \right)\right],\\
V^{(3)}&=& t_3\delta({\bf r}_i-{\bf r}_j)\delta({\bf r}_j-{\bf r}_k),
\end{eqnarray}
where $P_\sigma$ and $P_\tau$ denote the spin and isospin 
exchange operators, respectively.
The two-body part contains Wigner($w=1-m$), Bartlett($b$), 
Heisenberg($h$) and Majorana($m$) terms.
Concerning the spin-orbit force, the same form of the two-range Gaussian 
as the G3RS force \cite{LS}
is adopted. Coulomb force is approximated by the sum of seven Gaussians.

In the present work, we use the same interaction parameters 
as used in Ref.\cite{ENYObc}
except for $^{18}$O. Namely, we use the case 3 of MV1 force and choose the 
Bartlett, Heisenberg and Majorana parameters as $b=h=0$ and $m=0.576$,
respectively. The strengths of the spin-orbit force is chosen as 
$u_{I}=-u_{II}\equiv u_{ls}=900$ MeV.
For $^{18}$O, we can not obtain a stable solution of the AMD wave function
without parity
projection in case of the parameter $m=0.576$ due to a problem of the 
numerical calculation. It is because  
the Gaussian centers ${\bf Z}_i$ gather to the origin 
and the norm of the AMD wave function becomes almost zero
in the energy variation.
In order to avoid this problem, we use a slightly large 
Majorana parameter as $m=0.62$ instead of $m=0.576$.
We note that the properties of the ground state are not 
qualitatively unchanged in the parameter range
$m=0.576 \sim 0.63$ in most nuclei\cite{ENYOsup}.

\section{Results}
\label{sec:results}

We apply the present method of AMD to dipole excitations
in $^{8,10,14}$Be, $^{11,15,17}$B, $^{12,16,18,20}$C, $^{18}$O.
In the present AMD method without parity and spin projections, 
we can not obtain static solutions for the $N=8$ isotones 
by the cooling method 
due to the divergence of the inverse norm of the wave function, because
a system with $N=8$ favors the $p$-shell closed state, which is written by 
the AMD wave function with the zero-limit of Gaussian centers (${\bf Z}$) 
for all the neutrons.

\subsection{Properties of ground states}
The wave functions of the ground states($\Psi^0$) are obtained by the 
energy variation for the AMD wave function without spin-parity projections.
The width parameter $\nu$ is fixed and chosen to be an optimum value
for each nucleus to give the minimum energy of the ground state
in most cases.
For $^{15}$B, $^{16}$C, $^{18}$C, and $^{18}$O, 
we use a slightly larger width parameter than the optimum value to 
avoid the numerical problem in the norm of the wave function.
The adopted $\nu$ values are listed in Table.\ref{tab:nu}

\begin{table}
\caption{ \label{tab:nu} The adopted 
width parameters($\nu$) of the AMD wave functions.}
\begin{center}
\begin{tabular}{cccccccccccc}
& $^8$Be &$^{10}$Be & $^{14}$Be &	
$^{11}$B & $^{15}$B & $^{17}$B & $^{12}$C & $^{16}$C 
& $^{18}$C &$^{20}$C & $^{18}$O \\
$\nu$ (fm$^{-2}$) & 0.200&	
0.180&	0.175&	0.175&	0.180&	0.160&	0.180&	0.180&	0.175&	0.165&	0.170\\
\end{tabular}
\end{center}
\end{table}

\begin{figure}
\noindent
\epsfxsize=0.45\textwidth
\centerline{\epsffile{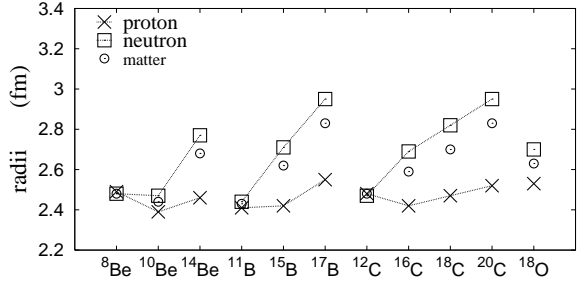}}
\caption{\label{fig:radii}
The root-mean-square radii of the ground states obtained by the AMD
calculation.
The radii for the proton(neutron) density distributions of the ground state,
$\Psi^0$, are plotted by crosses(squares). The circles indicate the nuclear 
matter radii.
}
\end{figure}

\begin{figure}
\noindent
\epsfxsize=0.45\textwidth
\centerline{\epsffile{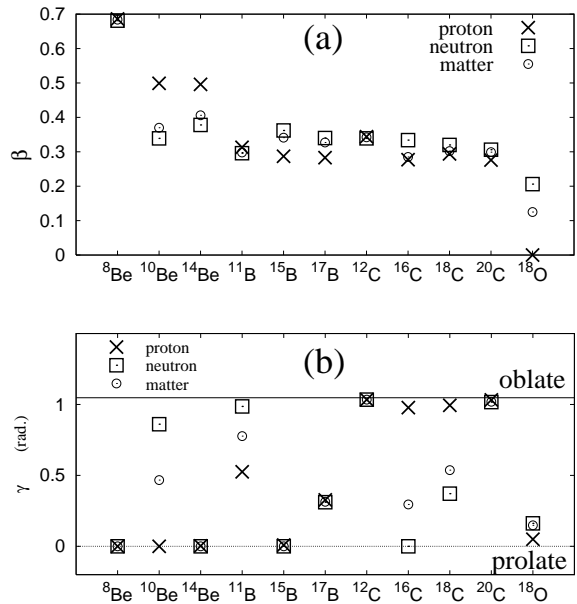}}
\caption{\label{fig:defo}
The deformations of the ground states calculated by AMD.
The deformation parameters (a)$\beta$ and (b)$\gamma$ for 
the proton(neutron) density distributions in $\Psi^0$ 
are plotted by crosses(squares). Circles are for the deformations of 
nuclear matter density.
}
\end{figure}

As is suggested in Refs.\cite{ENYObc,ENYO-c10,ENYO-c16,ENYOsup},
the shape of the proton and neutron density distribution rapidly changes
with the variation of the proton and neutron numbers.
The root-mean-square radii for the ground states($\Psi^0$) are shown in
Fig.\ref{fig:radii}. In each series of isotopes, 
the neutron radius is enhanced in the neutron-rich nuclei 
with the increase of neutron number. 
In Fig.\ref{fig:defo}, we show the deformation parameters ($\beta,\gamma$)
for proton and neutron densities. The results are the 
qualitatively same as the previous results obtained by the simple AMD
calculations \cite{ENYO-c10,ENYO-c16}. The difference of the shape 
between protons and neutrons is found in the $\gamma$ parameter 
as well as the $\beta$ value of some nuclei. 
The discrepancy of $\gamma$ is remarkable in $^{10}$Be and $^{16}$C. Namely, 
opposite deformation between proton and neutron densities appears in
these nuclei as discussed in Ref.\cite{ENYO-c10,ENYO-c16}.

\subsection{Energy weighted sum rule}
The energy weighted sum rule(EWSR) for isovector dipole resonances
is given by
\begin{equation} \label{eq:EWSR}
S(E1)=\int 
\hbar\omega \frac{dB(\omega;E1)}{d\omega} d\omega.
\end{equation} 
If the interaction commutes with the $E1$ operator, $S(E1)$ is identical
to the classical Thomas-Reiche-Kuhn(TRK) sum rule:
\begin{equation} 
S(TRK)= \frac{9 e^2}{8\pi M}\frac{NZ}{A},
\end{equation} 
where $M$ is the nucleon mass.
Due to the contributions of exchange terms and momentum dependent terms,
the interaction is usually incommutable with the $E1$ operator 
and $S(E1)$ is enhanced compared with $S(TRK)$.
Following the method explained in Ref.\cite{Kirson78},
we can calculate the EWSR 
with the initial-state expectation value of the double commutator of the 
Hamiltonian with the dipole operator $F$
\begin{equation}
S(E1)=\frac{1}{2} \langle  \Psi^0 |[F,[H,F]]|\Psi^0 \rangle.
\end{equation}
We estimate the enhancement factor, 
$\kappa=S(E1)/S(TRK)-1$, 
for the present interaction
by neglecting the contribution of the spin-orbit force.
The incommutable terms in the present interaction 
come from Heisenberg and Majorana exchange terms 
in the two-body central force $V^{(2)}$. 
If we write the two-body force as 
$V^{(2)}= v(r_{ij})+v^\tau (r_{ij})
\mbox{\boldmath$\tau_i$}\cdot \mbox{\boldmath$\tau_j$}$ 
($\mbox{\boldmath$\tau$}$ is the isospin 
SU(2) generator), the 
enhancement $\Delta S(E1)=S(E1)-S(TRK)$ is given as follows\cite{Kirson78},
\begin{equation}\label{eq:dels}
\Delta S(E1)=-\frac{3}{2\pi}e^2 \langle \Psi^0 |  
\sum _{i<j} r_{ij}^2 v^\tau(r_{ij})
\left( t_x(i)t_x(j)+t_y(i)t_y(j) \right)
   | \Psi^0 \rangle.
\end{equation}
Here $\mbox{\boldmath$t$}=\frac{1}{2}\mbox{\boldmath$\tau$}$.
By calculating the expectation value, Eq.(\ref{eq:dels}),
for the static solution $\Psi^0=\Phi({\bf Z}^0)$, we can obtain the 
values $S(E1)$ and $\kappa$.

In table \ref{tab:sum}, the calculated $S(E1)$ and $\kappa$
are shown.
In order to demonstrate that the sum rule is
kept in the present framework of the time-dependent AMD,
we compare the value $S(E1)=S(TRK)+\Delta S(E1)$ 
given by the static calculations with 
the EWSR value $S(E1;{\rm total})$ obtained 
by integrating the strengths Eq.(\ref{eq:EWSR}) 
calculated with the time-dependent AMD. 
As is shown in table \ref{tab:sum}, 
$S(E1)=S(E1;{\rm total})$ is practically satisfied. 
It is reasonable because the present calculation is regarded as
a method based on the small-amplitude TDHF. 
The enhancement factor $\kappa$ is 0.71 and 0.74 for $^{12}$C and $^{18}$O,
respectively, and it is $\kappa=0.6\sim 0.7$
for the neutron-rich Be, B, and C isotopes.
These are consistent with $\kappa=0.4 \sim 0.8$ due to the effect of 
the exchange mixtures of two-body interactions stated
 in Ref.\cite{RingSchuck}.
In the shell model calculations \cite{Sagawa-o,Sagawa-c}, 
the values $\kappa=40\%\sim50\%$ for $^{12}$C and $^{18}$O
are obtained for $S(E1;E_x< 40$ MeV) integrated up to 40 MeV,
while $\kappa=0.13$ for $S(E1;E_x<$ 40 MeV) in $^{18}$O is obtained by 
quasiparticle random phase approximation(QRPA)+phononcoupling model 
\cite{Colo}.
In the experimental photonuclear reactions, the 
observed cross section integrated up to 30 MeV for $^{12}$C\cite{Pywell} 
exhausts 63\% of TRK sum rule value,
and 90 \% of $S(TRK)$ is exhausted by EWSR integrated up to 42 MeV 
in $^{18}$O\cite{Woodworth79}.
As shown later, the EWSR is dominated by the GDR in the present results,
which means that the calculated GDR should be quenched and 
the large fraction of the strength should be in the higher 
energy region than the GDR.

\begin{table}
\caption{ \label{tab:sum} The energy weighted sum rule(EWSR) values of $E1$
transitions. The $S(E1;{\rm total})$ values 
are obtained by integrated the strengths up to
a enough large energy $E_x=100$ MeV in the time-dependent calculations 
of the AMD. The enhancement $\Delta S(E1)=S(E1)-S(TRK)$ is given by 
the ground state expectation values of the double commutator, 
Eq.(\protect\ref{eq:dels}). Here we ignore the contribution of the 
spin-orbit force.
The values of the enhancement factor $\kappa=S(E1)/S(TRK)-1$ are also shown.
The unit is MeV $e^2$ fm$^2$ for the energy weighted sum rule values.}
\begin{center}
\begin{tabular}{ccccccc}
 & $S(E1;{\rm total})$ & $S$(TRK) & $\Delta S(E1)$ & $S(E1)$ & 
$S(E1;{\rm total})$/$S(E1)$ & $\kappa$ \\ 
 &&&&&\%&
\\
\hline
$^{8}$Be	&	51 	&	30 	&	21 	&	51 	&	100 	&	0.71 \\
$^{10}$Be	&	58 	&	36 	&	23 	&	58 	&	99 	&	0.63 \\
$^{14}$Be	&	69 	&	42 	&	27 	&	69 	&	99 	&	0.62 \\
$^{11}$B	&	67 	&	40 	&	27 	&	67 	&	99 	&	0.66 \\
$^{15}$B	&	81 	&	49 	&	32 	&	81 	&	100 	&	0.63 \\
$^{17}$B	&	85 	&	52 	&	33 	&	86 	&	100 	&	0.63 \\
$^{12}$C	&	76 	&	44 	&	32 	&	76 	&	100 	&	0.71 \\
$^{16}$C	&	92 	&	56 	&	36 	&	92 	&	100 	&	0.65 \\
$^{18}$C	&	98 	&	59 	&	39 	&	98 	&	100 	&	0.65 \\
$^{20}$C	&	103 	&	62 	&	40 	&	102 	&	101 	&	0.66 \\
$^{18}$O	&	115 	&	66 	&	49 	&	115 	&	100 	&	0.74 \\
\end{tabular}
\end{center}
\end{table}

\subsection{Dipole resonances}
\subsubsection{$^{12}$C and $^{18}$O}
\begin{figure}
\noindent
\epsfxsize=0.35\textwidth
\centerline{\epsffile{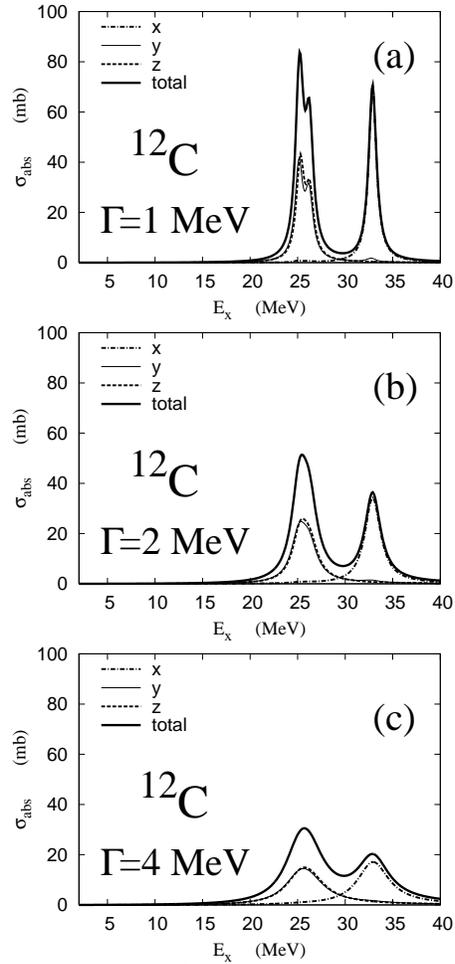}}
\caption{\label{fig:c12-s}
The calculated photonuclear cross section of $^{12}$C.
The smoothing parameter $\Gamma=1$, 2, 4 MeV are used.
Thin dash-dotted, solid, and dotted lines are the contribution of 
vibration for the $x$,$y$, and $z$-directions, respectively.
The thick solid lines indicate the total strengths.
}
\end{figure}

\begin{figure}
\noindent
\epsfxsize=0.35\textwidth
\centerline{\epsffile{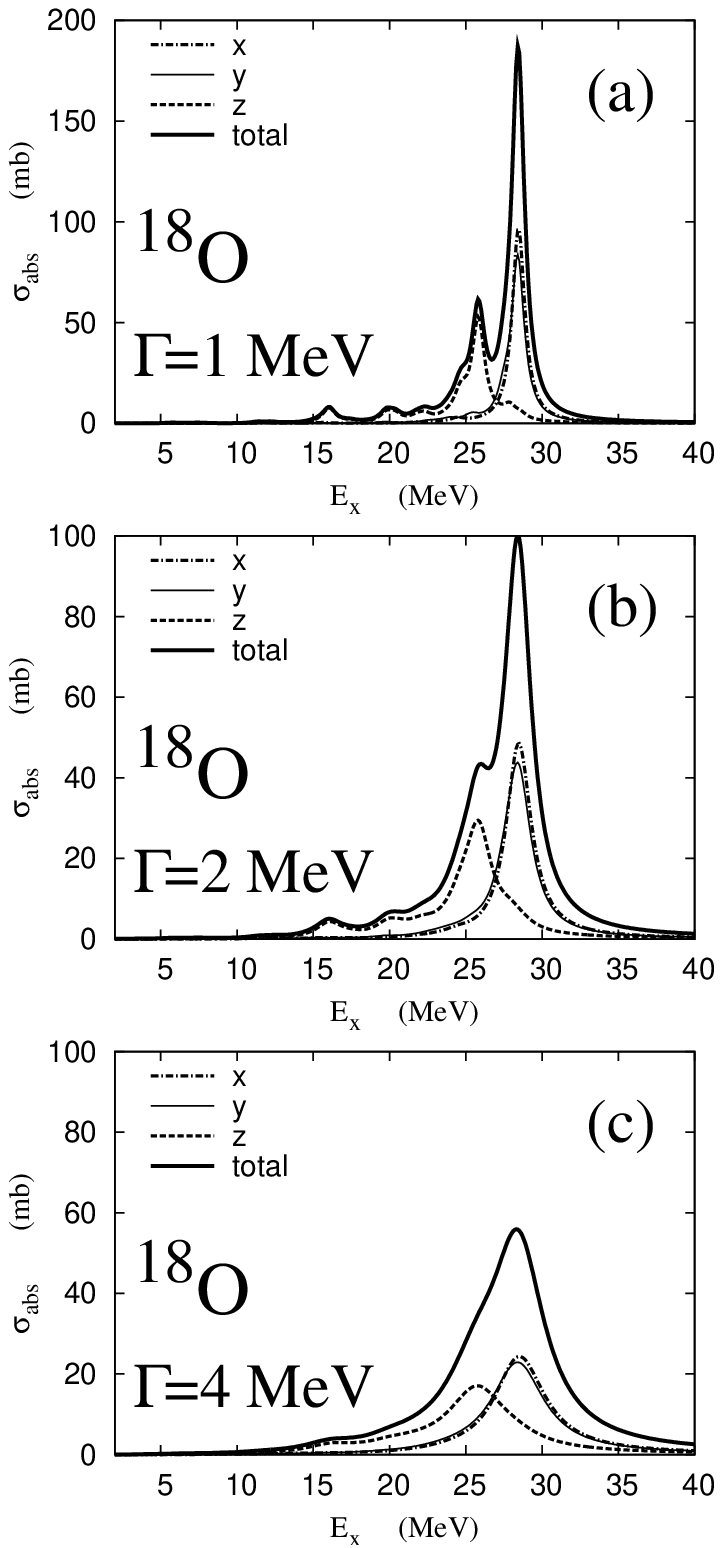}}
\caption{\label{fig:o18-s}
The calculated photonuclear cross section of $^{18}$O.
The smoothing parameter $\Gamma=1$, 2, 4 MeV are used.
Thin dash-dotted, solid, and dotted lines are the contribution of 
vibration for the $x$,$y$, and $z$-directions, respectively.
The thick solid lines indicate the total strengths.
}
\end{figure}

We first show the results of  the dipole resonances 
in $^{12}$C and $^{18}$O, 
and compare the results with other theoretical calculations and 
experimental data to see validity of the present method. 
The photonuclear cross section of $^{12}$C and $^{18}$O is plotted 
as a function of the excitation energy in Fig.\ref{fig:c12-s} and
\ref{fig:o18-s}.
Thin dash-dotted, solid, and dotted lines indicate the contribution of 
vibration for the $x$,$y$, and $z$-directions, respectively.
Here and hereafter, we chose the $x$, $y$, and $z$ axis as 
$\langle \Psi^0 |x^2 | \Psi^0 \rangle \le
\langle \Psi^0 |y^2 | \Psi^0 \rangle\le
\langle \Psi^0 |z^2 | \Psi^0 \rangle $ and 
$ \langle xy\rangle= \langle yz\rangle= \langle zx\rangle=0$.
The thick solid lines correspond to the total strengths.
We use the smoothing parameter $\Gamma=1,2,4$ MeV.
In the present results, the GDR peak lies at $E_x=26$ MeV and $E_x=28$ MeV
in $^{12}$C and $^{18}$O, respectively.
These peak positions are about 4 MeV higher than the observed GDR peaks 
\cite{Pywell,Woodworth79,Fultz,Berman76}, and also higher 
than other theoretical
values of the shell model\cite{Sagawa-o,Sagawa-c} and the 
QRPA calculations\cite{Colo}.
Compared with the observed photonuclear cross section, 
we need a smoothing parameter $\Gamma> 4$ MeV to 
reproduce the width of the GDR. The reason for such a large $\Gamma$
is considered to be due to the limitation of the present model space 
and lack of the effects of continuum states.
For the quantitative discussion of the magnitude of the GDR strength,
further quenching and the spreading are needed in the present calculations.

Although the quantitative description of the 
peak positions and the magnitudes of the GDR are not sufficient, 
the characteristic behavior of the calculated cross section is in 
reasonable agreement with that of the experimental data, and is consistent
with other theoretical calculations.
Since $^{12}$C has the oblate deformed ground state, 
the vibration for the $y$ and $z$-axes forms the GDR in the same energy, 
which results in an enhancement of the
lower part of the GDR. 
In the results of $^{18}$O, significant dipole strengths 
are distributed in the energy region
below the GDR due to the valence neutrons. 
The strong resonances have been experimentally 
observed in the region $10 - 15$ MeV\cite{Leistenschneider,Woodworth79}, 
and about 8\% 
\cite{Leistenschneider} of the TRK sum rule is exhausted
 by the integrated strength 
of the experimental data up to $E_x=15$ MeV.
These low-lying resonances 
are well described by the shell model calculations \cite{Sagawa-o}, which 
gives $S(E1;E_x<15{\rm MeV})/S(TRK)=6\%$.
In the present results, the excitation energies of these 
low-lying resonances seem to be overestimated compared with 
those of the shell model and QRPA calculations as well as the GDR. 
Namely, the strengths of the low-lying peaks are distributed in the
$E_x=15 \sim 20$ region, and $S(E1;E_x< 17$ MeV)/$S(TRK)=6\%$
and $S(E1;E_x< 15$MeV)/$S(TRK)=3\%$ in the present results.
Considering the shift of the excitation 
energies, we can state that the calculated strengths of the low-lying 
resonances reasonably agree to the experimental data.

One of the reason for the overestimating excitation energies 
of the dipole resonances is considered to be because the surface 
diffuseness may be 
underestimated by the simple AMD wave function. It might be improved by
the extension of the model wave function such as the deformed base AMD
proposed by one(M.K.) of the authors and his collaborators \cite{Kimura01}. 

\subsubsection{C, B and Be isotopes}
\begin{figure}
\noindent
\epsfxsize=0.35\textwidth
\centerline{\epsffile{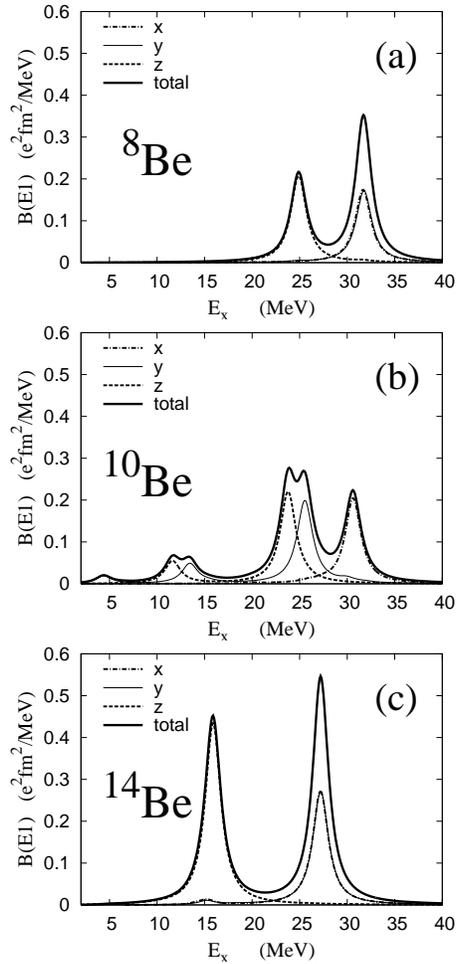}}
\caption{\label{fig:be-iso}
The calculated $E1$ transition strengths
of $^{8}$Be, $^{10}$Be, and $^{14}$Be.
The smoothing parameter is chosen to be $\Gamma=2$ MeV.
Thin dash-dotted, solid, and dotted lines are the contribution of 
vibration for the $x$,$y$, and $z$-directions, respectively.
The total strengths are shown by the thick solid lines.
}
\end{figure}

\begin{figure}
\noindent
\epsfxsize=0.75\textwidth
\centerline{\epsffile{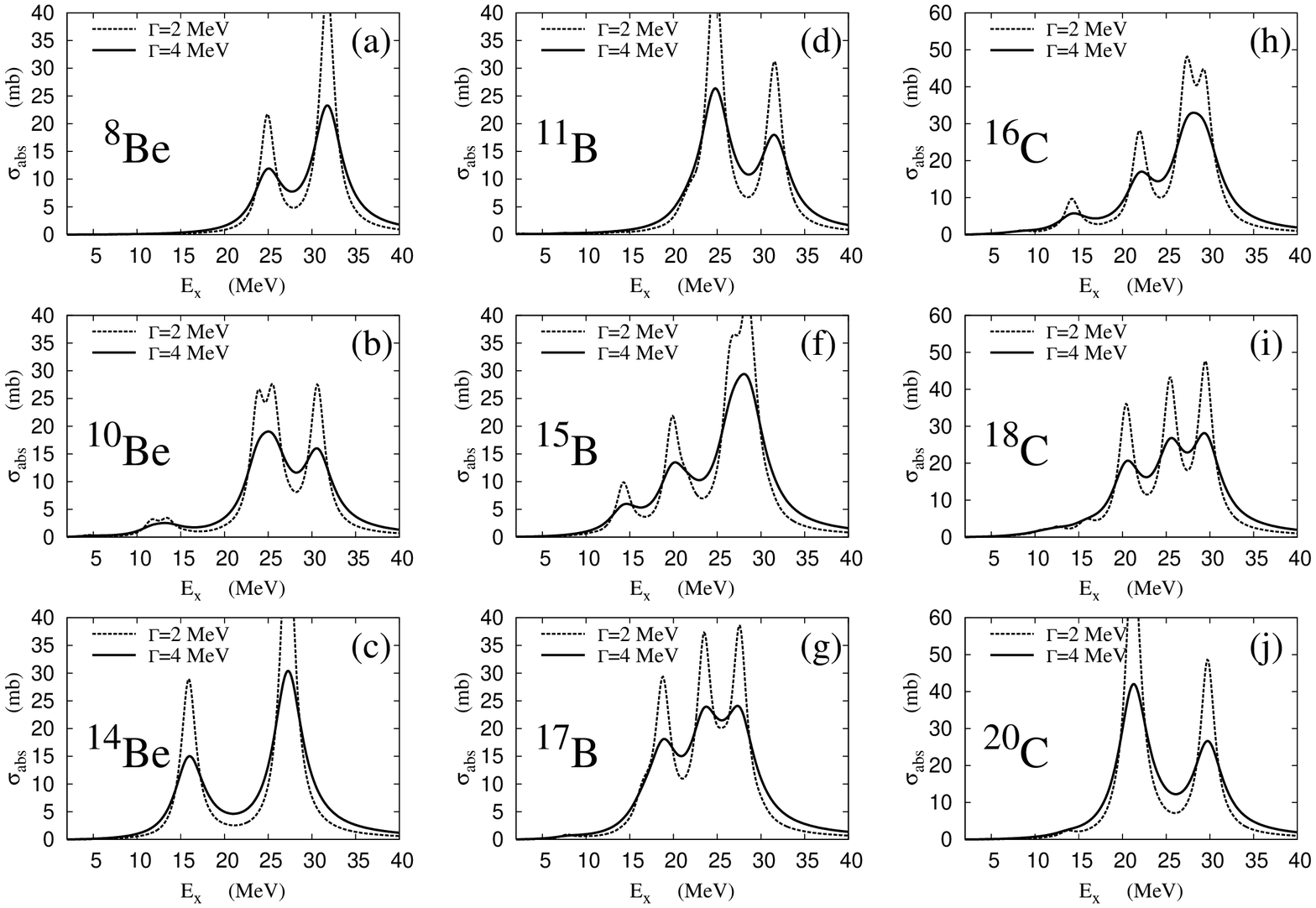}}
\caption{\label{fig:s-abs}
The calculated photonuclear cross section in
(a)$^8$Be, (b)$^{10}$Be, (c)$^{14}$Be,
(e)$^{11}$B, (f)$^{15}$B, (g)$^{17}$B,
(h)$^{16}$C, (i)$^{18}$C, (j)$^{20}$C,
The smoothing parameter is chosen to be $\Gamma=2$ MeV(dashed lines) and 
4 MeV(solid lines).
}
\end{figure}

Next we investigate the dipole resonances in 
neutron-rich Be, B, and C isotopes. 
We show the $E1$ strengths in $^{8}$Be, $^{10}$Be, and $^{14}$Be 
in Fig.\ref{fig:be-iso}, and the photonuclear 
cross section in the left column of Fig.\ref{fig:s-abs}.
In $^{10}$Be, 
the dipole resonances in $^{10}$Be can be decomposed into two parts 
 $E_x< 15$ MeV and $E_x > 20$ MeV. 
The former consists of the soft resonances with the dominant strengths in 
$10< E_x< 15$ MeV. 
The latter contains the GDR with a double peak structure with 
$7- 8$ MeV energy splitting, which is
similar to that of $^{8}$Be. 
In the $E_x<5$ MeV region, we find a peak with the strength $B(E1)=0.06$
$e^2$fm$^2$. We consider this is a $1^-$ state and corresponds to the 
known $1^-$ state at 5.96 MeV.
The present low-lying peaks in the  $10< E_x< 15$ MeV
originate in the cluster structure. The details will be discussed in the 
next section.
In the TDHF+absorbing boundary condition(ABC)
calculations, there are not
such the significant strengths of the soft $E1$ resonances
\cite{Nakatsukasa}.
On the other hand, the GDR of the TDHF+ABC 
calculations is consisitent with the present results.
In $^{14}$Be, the GDR splits into two peaks at $E_x=15$ MeV and at
$E_x> 25$ MeV. The lower peak appears in the vibration 
along the longitudinal axis($z$). 
As seen in Figs.\ref{fig:radii} and \ref{fig:defo}, 
$^{14}$Be has the large prolate 
deformation of the neutron density as well as the large neutron radius,
and hence, it has the enhanced neutron skin structure 
along the longitudinal direction.
The decrease of the excitation energy of the lower
GDR peak is naturally understood because of the developed neutron skin.
Also in the TDHF+ABC calculations \cite{Nakatsukasa}, the GDR for the 
longitudinal motion appears at $E_x=15$ MeV, while the higher peak 
for the transverse motion is around $E_x=25$ MeV. 
Although the peak position of the GDR for each direction 
is similar to the present results,
the GDR is not splitting in the TDHF+ABC results 
because the widths are largely spread. Another difference with the 
present results is that there exists 
a very soft resonance at $E_x\sim 5$ MeV in TDHF+ABC.
These differences seem natural because of the following reason.
The spreading and quenching of the GDR
may be large in $^{14}$Be, which has a small neutron 
separation energy and therefore the dipole strengths should 
be affected by the continuum states and the long tail of the 
neutron halo structure. These effects are not taken into account 
in the present framework, while they are included in the TDHF+ABC. 
In the shell model calculations, the large $E1$ strength is found in 
the low energy region ($5< E_x <15$ MeV) when including 
the $fp$ shell configurations with a Warburton-Brown(WBP) interaction
\cite{Sagawa-be}.

\begin{figure}
\noindent
\epsfxsize=0.35\textwidth
\centerline{\epsffile{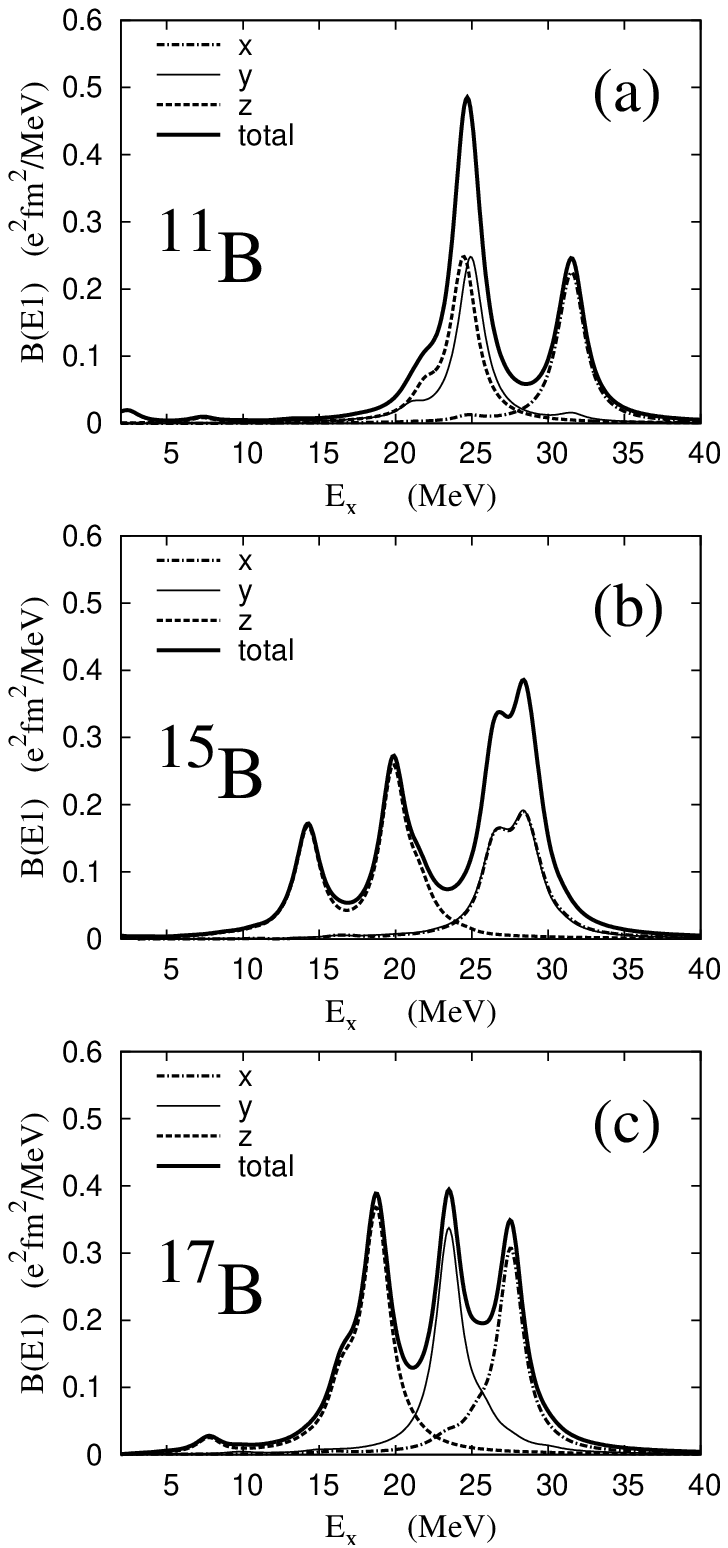}}
\caption{\label{fig:b-iso}
The calculated $E1$ transition strengths
of $^{11}$B, $^{15}$B, and $^{17}$B.
The smoothing parameter is chosen to be $\Gamma=2$ MeV.
Thin dash-dotted, solid, and dotted lines are the contribution of 
vibration for the $x$,$y$, and $z$-directions, respectively.
The total strengths are shown by the thick solid lines.
}
\end{figure}

The $E1$ strengths and the photonuclear cross section in B isotopes
are shown in Figs.\ref{fig:b-iso} and \ref{fig:s-abs}.
In the B isotopes, the feature of the GDR changes reflecting
the variation of the deformation as the neutron number increases. 
In $^{15}$B, two peaks 
of the GDR appear at $E_x=20$ and 27 MeV
(see Fig.\ref{fig:b-iso}(b)). Due to the prolate deformation,
the lower GDR at $E_x=20$ for the longitudinal($z$-direction) vibration 
has a smaller strength than that of the higher one. 
The excitation energy of the
GDR is smaller than that of the $^{11}$B.
One of the unique features in $^{15}$B is that
a soft resonance appears in $E_x=10- 17$ MeV region, which 
exhausts $16\%$ of the $S(TRK)$ value. 
This soft resonance arises from the longitudinal vibration and 
decouples energetically with the GDR region($E_x>17$) MeV.
In $^{17}$B, the GDR peaks spread over a wide energy region
due to the triaxial deformation. The peak position of the lowest GDR further 
shifts toward the low energy region: $E_x\sim 18$ MeV. We can not find 
strong soft $E1$ resonance other than the GDR in $^{17}$B.

\begin{figure}
\noindent
\epsfxsize=0.35\textwidth
\centerline{\epsffile{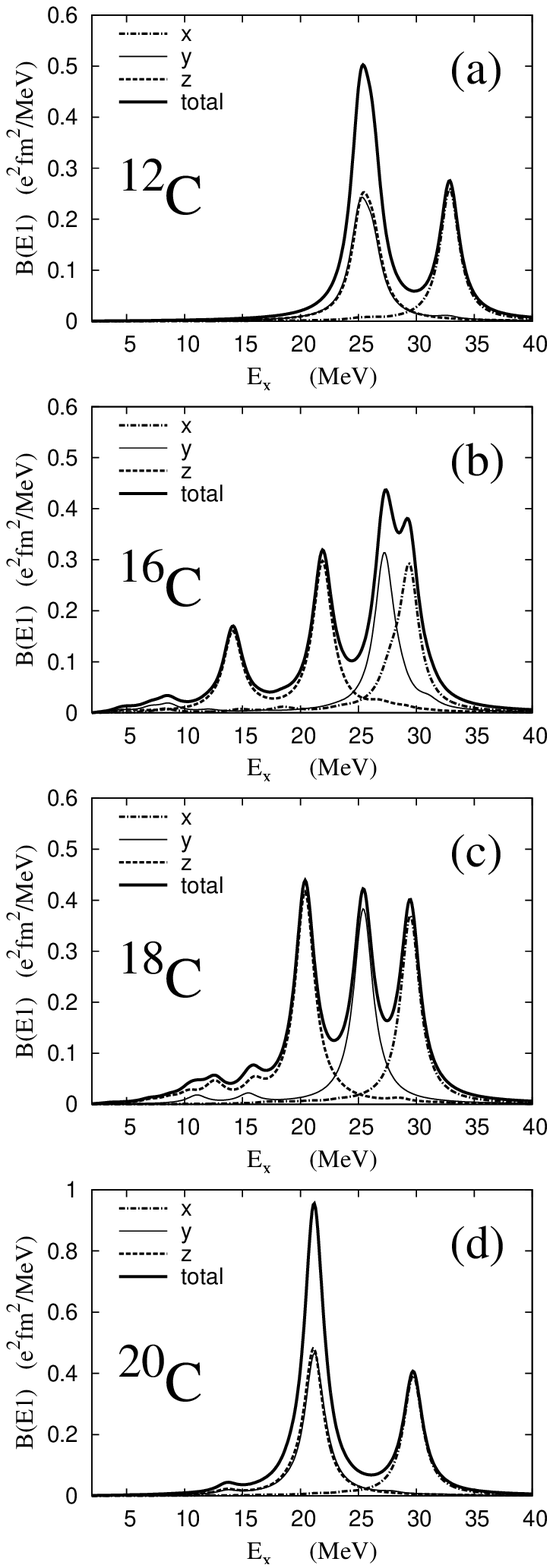}}
\caption{\label{fig:c-iso}
The calculated $E1$ transition strengths
of $^{12}$C, $^{16}$C, $^{18}$C and $^{20}$C.
The smoothing parameter is chosen to be $\Gamma=2$ MeV.
Thin dash-dotted, solid, and dotted lines are the contribution of 
vibration for the $x$,$y$, and $z$-directions, respectively.
The total strengths are shown by the thick solid lines.
}
\end{figure}

We show the results of C isotopes in Figs.\ref{fig:c-iso} and \ref{fig:s-abs}. 
In comparison between B(Fig.\ref{fig:b-iso}) 
and C isotopes(Fig.\ref{fig:c-iso}), it is found that 
the feature of the dipole transitions in $^{16}$C is quite similar 
to that in $^{15}$B, which has the same neutron number($N=10$) with 
$^{16}$C. Namely, the dipole strength for the longitudinal($z$-axis)
 vibration splits into two peaks, the GDR at $E_x=22$ MeV and a 
soft resonance at $E_x=14$ MeV.
As a result, $^{16}$C has a significant dipole strength in the
low-energy region($E_x<17$ MeV) below the GDR region. This soft dipole
resonance at $E_x=14$ MeV is consistent with 
the shell model calculations\cite{Sagawa-c},
where a remarkable peak is found at the same energy.
In $^{18}$C,
since it has a triaxial deformation as well as $^{17}$B, 
the shape of the strength function in the GDR region
is similar to that of $^{17}$B, though the peak positions are slightly higher
than those of $^{17}$B. A difference between $^{18}$C
and $^{17}$B is the soft dipole strengths 
in the energy region $E_x<17$  MeV.
Although there is no noticeable
peak in this energy region, we find some fractions of the dipole 
strengths in $^{18}$C rather than in $^{17}$B.
In $^{20}$C with an oblate deformation, 
the shape of the strength function $dB(E1,\omega)/d\omega$ is 
similar to the $^{12}$C, while the peak positions are 4-5 MeV lower than
those of $^{12}$C.

\section{Discussion}\label{sec:discuss}

As shown before, the
remarkable peaks of the soft resonances
are found in the dipole strengths of $^{10}$Be, $^{15}$B and $^{16}$C. 
They are separated clearly from the GDR region.
It is natural to expect that 
these soft resonances arise from the coherent excitations 
of excess neutrons. In order to link the $E1$ resonances with 
collective motions we analyze the time evolution of the single-particle 
wave functions. In the time-dependent AMD, the expectation value 
of the dipole operator ${\cal M}(E1,\mu=0)$ for the $\Psi(t)=\Phi({\bf Z(t)})$
is directly related to the real part of the centers of the 
single-particle Gaussian wave packets
\begin{equation}\label{eq:zcomponent}
\langle \Psi(t)|{\cal M}(E1,\mu=0)|\Psi(t)\rangle
=\frac{N}{A}\sum_{i=1}^Z {\rm Re}\left[\frac{Z_{iz}(t)}{\sqrt{\nu}}\right]
-\frac{Z}{A}\sum_{i=Z+1}^A {\rm Re}\left[\frac{Z_{iz}(t)}{\sqrt{\nu}}\right],
\end{equation}
where the $i$-th particle is a proton(neutron) for 
$i=1,\cdots, Z(Z+1,\cdots,A)$, and $Z_{iz}$ is the $z$-component of 
the center ${\bf Z}_i$ for the $i$-th single particle Gaussian wave function.
It should be stressed that 
the $E1$ excitations are expressed by the motion of 
the centers of single-particle Gaussian wave packets. Since 
the $E1$ strength is given by the Fourier transform of 
Eq.(\ref{eq:zcomponent}) as explained in (\ref{sec:formulation}),
we can examine 
separately the contribution of the motion of each single-particle wave packet
to the dipole 
strengths by Fourier transform of ${\rm Re}\left[Z_{iz}(t)/\sqrt{\nu}\right]$
and explicitly see collective modes. As discussed later, 
in case that a collective mode due to the inter-cluster motion appears,
the mode can be seen as a peak at the corresponding excitation energy 
in the sum of the components for 
nucleons in each cluster,
\begin{equation}\label{eq:dz}
-\frac{1}{\pi\epsilon} {\rm Im}\int dt 
\sum_{i\in C_k} {\rm Re}[Z_{iz}/\sqrt{\nu}] 
e^{i\omega t},
\end{equation}
where $C_1,C_2,\cdots$ are the constituent clusters, and
$\epsilon$ is the same parameter in Eq.(\protect\ref{eq:BE1}).

\begin{figure}
\noindent
\epsfxsize=0.6\textwidth
\centerline{\epsffile{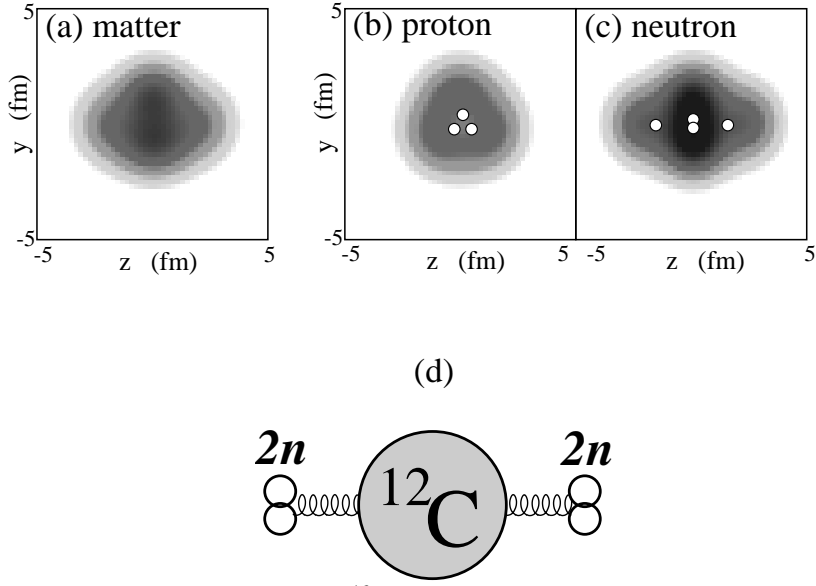}}
\caption{\label{fig:c16dense}
The density distribution of the ground state of $^{16}$C.
The densities are integrated along the $x$-axis.
(a)Matter density, (b) proton density, (c) neutron density 
are shown. The spatial configuration of the centers of single-particle
Gaussian wave packets for protons and neutrons are also plotted 
by open circles in (b) and (c), respectively. 
Each circle consists of a spin-up nucleon and a spin-down nucleon.
(d): The schematic figure for the configurations of the Gaussian centers
in $^{16}$C written by $2n+^{12}$C+$2n$. 
}
\end{figure}

\begin{figure}
\noindent
\epsfxsize=0.36\textwidth
\centerline{\epsffile{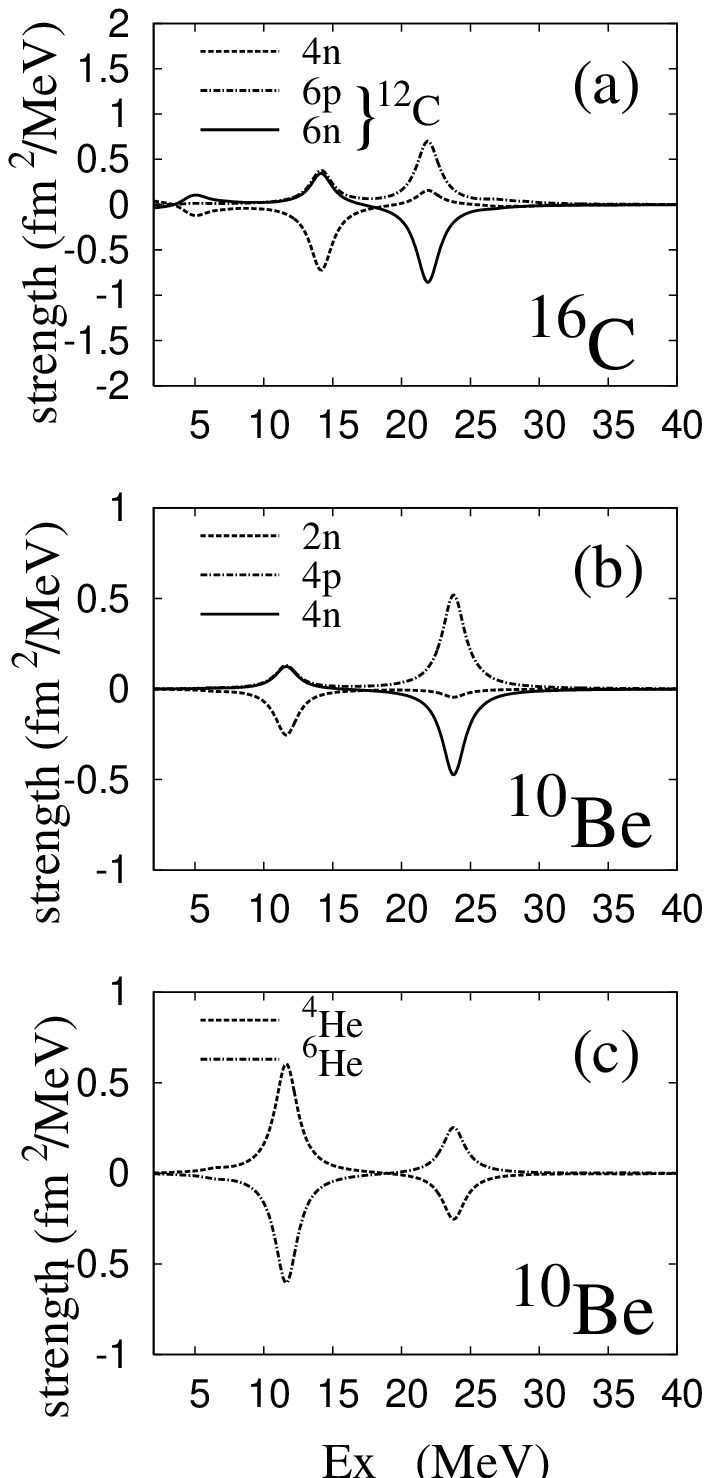}}
\caption{\label{fig:dz}
Strength of the motion of the Gaussian centers
in the response of $^{16}$C and
$^{10}$Be with an instantaneous dipole field along $z$-axis. 
The strength is given by Fourier transform of positions of the Gaussian centers
as shown in Eq.(\protect\ref{eq:dz}). 
We divide $A$ nucleons into some groups: $C_1,C_2,\cdots$ and 
sum up the components for nucleons in each group.
(a)Motion of the Gaussian centers in $^{16}$C. $C_1$ consists of 
four valence neutrons, and $C_2$($C_3$) contains 6 protons(neutrons) in 
the $^{12}$C core. 
(b)Motion in $^{10}$Be. $C_1$, $C_2$ and $C_3$ consist of 
two valence neutrons, 4 protons, and 4 residual neutrons.
(c)Relative motion between $^{4}$He($C_1$) and $^{6}$He($C_2$) 
clusters in $^{10}$Be. The dotted, dash-dotted, and solid lines correspond
to $C_1$, $C_2$, and $C_3$.}
\end{figure}

In Fig.\ref{fig:c16dense}, 
we illustrate the density distribution and the spatial 
configuration of the Gaussian centers 
${\rm Re}[{\bf Z}_i/\sqrt{\nu}]$ in the static solution of $^{16}$C.
There is the difference between proton and neutron densities in the ground state.
The $E1$ transitions is described by the small-amplitude motion around this
static solution.
We see a $2n$+$^{12}$C+$2n$ configuration in the spacial configuration of the 
Gaussian 
centers, which forms the prolate neutron deformation with the longitudinal $z$ 
axis. After the instantaneous external dipole field ${\cal M}(E1,\mu=0)$ is
imposed, four valence neutrons coherently move 
against the core $^{12}$C with
the oscillation energy $E_x\sim 14$ MeV to form the soft dipole peak.
On the other hand, it is found that the strengths in the GDR region
($E_x>20$ MeV) arise from the motion of the nucleons 
within the $^{12}$C core. In figure \ref{fig:dz}(a),
we show the strengths of the motion for 4 valence neutrons, 6 protons and
6 neutrons in the $^{12}$C core. It is found that 
the valence neutrons move with a negative strengths against 6 
protons and 6 neutrons in the $E_x\sim 14$ MeV region, while, in the GDR region, 
the dipole strength is dominated by the relative motion between 6 protons and 
6 neutrons inside the $^{12}$C core. 
The reason why four neutrons move coherently in the 
$2n$+$^{12}$C+$2n$ configuration is easily understood as follows.
Since the $2n$+$^{12}$C+$2n$ configuration is linear,
one can consider a configuration with two dineutrons on the opposite
sides of the core($^{12}$C).
Let us imagine the inert three clusters($2n$+$^{12}$C+$2n$), 
which are connected with
two identical springs as shown in Fig.\ref{fig:c16dense}(d).  
In the motion along the longitudinal axis, 
there are two eigen modes of the oscillation.
One is the mode where two dineutrons move in phase and
the core moves in the opposite way, and
the other is the one with the opposite motion of the dineutrons to each other.
The former corresponds to the isovector dipole mode. Thus, the coherent 
motion of the valence neutron can be interpreted by the relative 
motion in the $2n$+$^{12}$C+$2n$ configuration.
The soft dipole resonance due to the excess neutrons in the present result
 of $^{16}$C
well corresponds to the shell model calculations\cite{Sagawa-o}, where the 
$0p\rightarrow 1s0d$ and 
$1s0d\rightarrow 0f1p$ transitions work coherently to enhance the strength 
at $E_x=12-14$ MeV in $^{16}$C.

\begin{figure}
\noindent
\epsfxsize=0.6\textwidth
\centerline{\epsffile{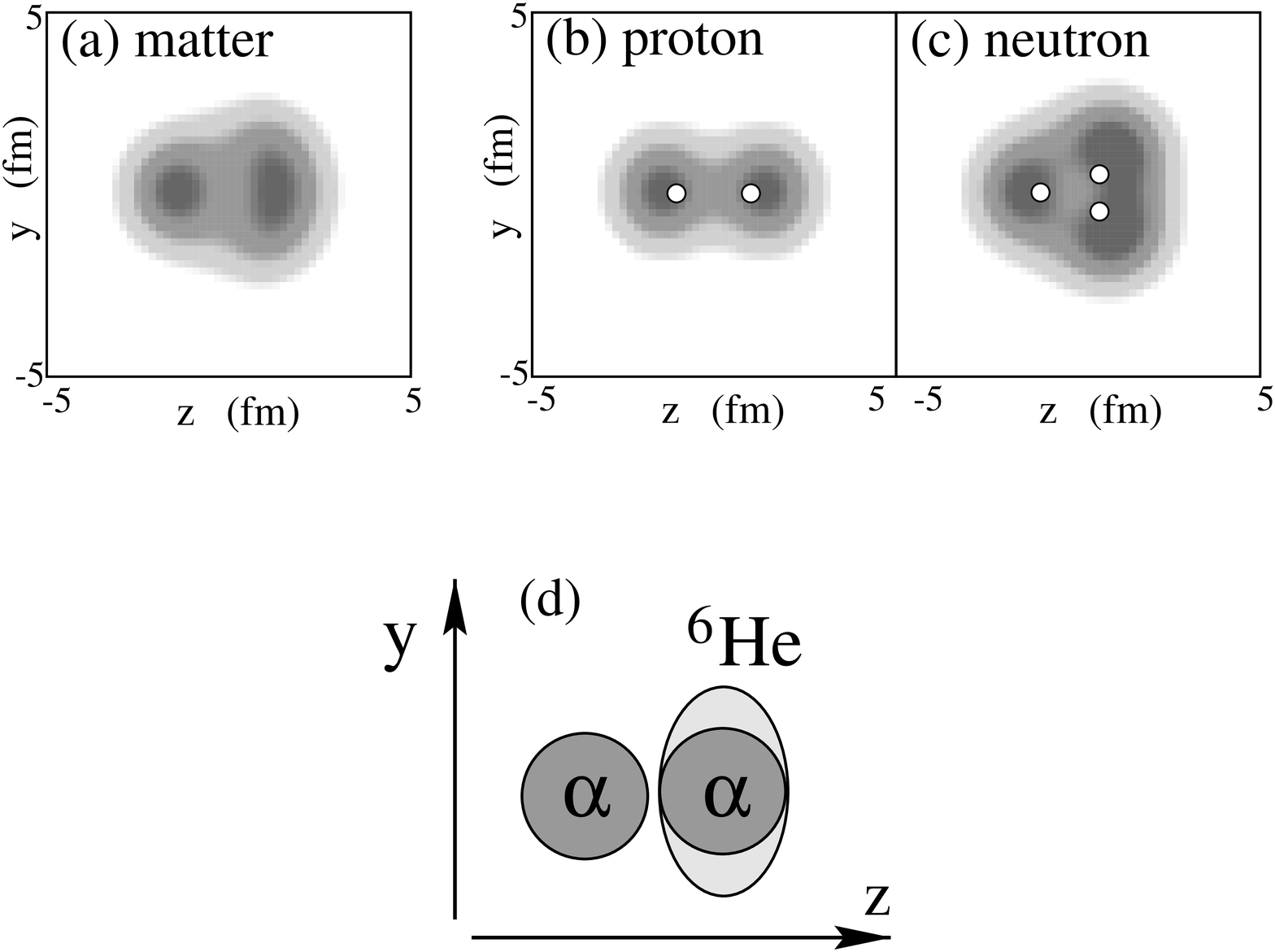}}
\caption{\label{fig:be10dense}
The density distribution of the ground state of $^{10}$Be.
The densities are integrated along the $x$-axis.
(a)Matter density, (b) proton density, (c) neutron density 
are shown. The spatial configuration of the centers of single-particle
Gaussian wave packets for protons and neutrons are also plotted 
by open circles in (b) and (c), respectively. 
Each circle consists of a spin-up nucleon and a spin-down nucleon.
(d): The schematic figure for the structure of $^{10}$Be with
$\alpha+^6$He configuration.
}
\end{figure}

We show the density distribution of the ground state of $^{10}$Be in Fig.
\ref{fig:be10dense}. In order to understand collective motion for 
the soft dipole resonances, it is useful to regard the $^{10}$Be as 
the $\alpha+^6$He cluster state. In the analysis of 
motion of the Gaussian centers, it is 
found that the strength at $E_x=10-15$ MeV contains two independent modes.
One is the inter-cluster motion between $\alpha$ and $^6$He, which contributes
to the resonance at $E_x=12$ MeV in the longitudinal vibration 
along the $z$-axis. The other is the coherent motion of the valence neutron 
against the core $^8$Be, which results in the resonance at $E_x=14$ MeV
in the vibration along the $y$-axis.
In Fig.\ref{fig:dz}(c), we show the strength
of the $\alpha$-$^6$He inter-cluster motion. It has a dominant peak 
at $E_x=10-15$ MeV, which corresponds to the soft peak in the
$z$-component.
 On the other hand, the GDR is described by
the motion inside the core $^8$Be.
Also in $^{15}$B, we find that 
the coherent motion of the valence neutrons 
contributes to enhance the strengths of the soft resonance.
It is concluded that the remarkable peaks at $E_x=10-15$ MeV 
in $^{16}$C, $^{10}$Be and $^{15}$B arise from  
the coherent motion of the valence nucleons, which decouples
with the motion inside 
the core.

We show the calculated photonuclear cross section 
in Fig.\ref{fig:s-abs}. 
The shape of the strength function in the GDR region($E_x>17$ MeV) has a close 
relation with the deformation of the system.
In the oblately deformed system such as $^{11}$B and $^{20}$C as well 
as $^{12}$C, the GDR splits in two part. The lower GDR peak has  
large transition strength. In the neutron-rich nucleus, $^{20}$C,
the excitation energy of the GDR 
is the lowest among the three nuclei.
Also in the $^{8}$Be and $^{14}$Be with the prolate deformation, 
the GDR splits into two, but the higher GDR has larger strength
in contrast to the oblate deformation.
The lower resonance in $^{14}$Be exists at $E_x=15$ MeV.
In triaxial nuclei such as $^{17}$B and $^{18}$C, the GDR consists of 
three peaks with 5 MeV of the energy splitting. However, considering 
spreading of the widths, three peaks may overlap to form a
broad structure in the $E_x=15-30$ MeV.
$^{10}$Be and $^{16}$C have the ground state 
with opposite deformations between proton and neutron
densities. In spite of this unusual properties of the deformations, 
we can not find an abnormal feature in the dipole strength of the 
GDR region in these nuclei. 
As explained before, the remarkable soft peaks appear at $E_x=10-15$ MeV 
due to the motion of the valence nucleons against to the core, while
the GDR at $E_x>17$ MeV arises from the vibration within the core.
Therefore it is considered that the GDR may reflect mainly 
the features of the core nuclei instead of the 
deformations of the total system. In fact, the GDR of $^{10}$Be 
lies at a similar excitation energy to that of $^8$Be.

\begin{figure}
\noindent
\epsfxsize=0.35\textwidth
\centerline{\epsffile{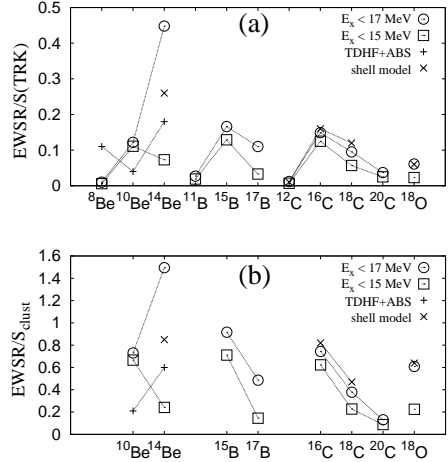}}
\caption{\label{fig:sum}
The ratio of the EWSR integrated for the soft dipole resonances
to $S(TRK)$ and $S_{\rm clust}$. 
Circles indicate the ratio for $S(E1;E_x < 17$ MeV) and 
squares are for $S(E1;E_x<15$ MeV). 
The smoothing parameter is chosen to be $\Gamma=2.0$ MeV.
The results of TDHF+ABC for 
$S(E1;E_x<15$ MeV) in Be isotopes are plotted by the symbols ``$+$''.
The symbols ``$\times$'' show  the results of the shell model calculations 
for $S(E1;5< E_x <15$ MeV), $S(E1; E_x <14$ MeV), $S(E1; E_x <15$ MeV)  
in Be\protect\cite{Sagawa-be}, 
C\protect\cite{Sagawa-c}, and O\protect\cite{Sagawa-o}, respectively.
}
\end{figure}

Below the GDR region, we find the significant strengths of the soft resonances 
in such neutron-rich 
nuclei as $^{10}$Be, $^{15}$B, $^{16}$C and $^{18}$C 
due to the excess neutrons.
Especially, $^{10}$Be, $^{15}$B and $^{16}$C have the
remarkable soft peaks, which are decoupled with the GDR.
  In the neutron-rich nuclei, the cluster 
sum rule\cite{Alhassid} is a convenient 
measure to estimate the contribution of the 
motion of the excess neutrons in the dipole strengths\cite{Sagawa90}.
Assuming clustering with a core and $N_v$ valence neutrons, we consider
the core cluster with $Z_1=Z, N_1=N-N_v$ and the valence cluster with
$Z_2=0, N_2=N_v$.  The cluster sum rule is given as,
\begin{equation}
S_{\rm clust}=\frac{\hbar^2}{2m}\frac{9}{4\pi}
\frac{(Z_1 N_2-Z_2 N_1)^2}{A(Z_1+N_1)(Z_2+N_2)}e^2
=\frac{\hbar^2}{2m}\frac{9}{4\pi}
\frac{Z^2N_v^2}{A(A-N_v)N_v}e^2.
\end{equation}
This value is the remainder when one subtract the core contribution of the 
classical EWSR from the total $S(TRK)$ value. Consequently, 
$S_{\rm clust}$ is the margin which indicates the contribution 
of the excess neutrons.
The integrated strength of the low-energy resonances 
should be compared with $S_{\rm clust}$ to see
the softness and collectivity of the resonances due to the valence neutron
motion against the core.
In table \ref{tab:c-sum}, the EWSR for the low-lying resonances
are listed with the values of the classical TRK sum rule
and the cluster sum rule($S_{\rm clust}$).
In the derivation of the cluster sum rule $S_{\rm clust}$, the core cluster 
is assumed to be $^{8}$Be, $^{11}$B, $^{12}$C and $^{16}$O,
in Be, B, C and O isotopes, respectively.
We show the EWSR values integrated up to $E_x<15$ MeV 
and $E_x<17$ MeV in the present results, and the EWSR with other calculations.
The ratios of the EWSR to $S(TRK)$ and $S_{\rm clust}$ are shown in 
Figs.\ref{fig:sum}(a) and \ref{fig:sum}(b), respectively.
In C isotopes, 
$S(E1;E_x< 17$ MeV) is the largest in $^{16}$C
and it declines in further neutron-rich C isotopes.
The present results of C isotopes well agree to 
the shell model calculations\cite{Sagawa-o}. 
Also in B isotopes, the similar feature is found. Namely,
$S(E1;E_x< 17$ MeV) is the largest in $^{15}$B
and it decreases in further neutron-rich nucleus $^{17}$B.
The striking point is that 
the EWSR for the low-lying resonances is remarkably enhanced 
in the moderately neutron-rich nuclei 
with an appropriate number of excess neutrons,
but it is suppressed in very neutron-rich nuclei.
It is reasonable because the enhancement of the soft dipole strengths
is due to the coherent motion of the valence neutrons relative to 
the core. It means that the decoupled collective 
modes appear based on the relative motion between the core and 
valence neutrons and the motion inside the core, in $^{15}$B and $^{16}$C. 
On the other hand, 
as the neutron skin develops in further neutron-rich nuclei $N>10$, 
the motion of the excess neutrons is
not decoupled but they join the neutrons
 inside the core. As a result, in $^{17}$B and $^{20}$C, 
the soft dipole mode is assimilated into the GDR, 
and the excitation energy of the GDR decreases.
Also in $^{10}$Be, the EWSR for the low-lying resonances is significant
as well as $^{15}$B and $^{16}$C. In these nuclei, 
the cluster sum rule value $S_{\rm clust}$
is almost exhausted by the calculated $S(E1;E_x<17$ MeV).
In $^{14}$Be, the $S(E1;E_x < 17$ MeV) is very large 
due to the peak at $E_x=16$ MeV. 
The reason for the enhanced $S(E1;E_x<17$ MeV) in $^{14}$Be is
different from other nuclei($^{10}$Be,$^{15}$B and $^{16}$C). 
In case of $^{14}$Be, 
the enhanced $S(E1;E_x<17$ MeV) does not originate in the soft 
resonance decoupled from 
the GDR, but the GDR for the longitudinal vibration 
itself contributes the EWSR for low-energy region
because it becomes soft due to the
prolate deformation with a developed neutron skin structure.

\begin{table}
\caption{ \label{tab:c-sum} The Energy-weighted sum rule 
for the soft dipole resonances.
The energy weighted sum integrated up to $E_x < 17 $ MeV and
$E_x < 15$ MeV are shown in the fourth and fifth columns.
The core cluster in the derivation of the cluster sum rule $S_{\rm clust}$
is $^{8}$Be, $^{11}$B, $^{12}$C and $^{16}$O,
in Be, B, C and O isotopes, respectively.
The smoothing parameter is chosen to be $\Gamma=2.0$ MeV.
The unit is $e^2$fm$^2$MeV.}
\begin{center}
\begin{tabular}{ccccc}
 & $S(TRK)$ & $S_{\rm clust}$ & \multicolumn{2}{c}{$S(E1)$} \\
 &          &                 & $E_x < 17$ MeV       & $E_x <15$ MeV \\
\\
\hline
$^{8}$Be	&	29.7 	&	$-$	&	0.3 	&	0.2 \\
$^{10}$Be	&	35.6 	&	5.9 	&	4.3 	&	4.0 \\
$^{14}$Be	&	42.4 	&	12.7 	&	19.0 	&	3.1 \\
$^{12}$C	&	44.5 	&	$-$	&	0.5 	&	0.3 \\
$^{16}$C	&	55.6 	&	11.1 	&	8.3 	&	6.9 \\
$^{18}$C	&	59.3 	&	14.8 	&	5.6 	&	3.4 \\
$^{20}$C	&	62.3 	&	17.8 	&	2.3 	&	1.6 \\
$^{11}$B	&	40.5 	&	$-$	&	1.1 	&	0.8 \\
$^{15}$B	&	49.4 	&	9.0 	&	8.2 	&	6.4 \\
$^{17}$B	&	52.3 	&	11.9 	&	5.8 	&	1.7 \\
$^{18}$O	&	65.9 	&	6.6 	&	4.0 	&	1.5 \\
\end{tabular}
\end{center}
\end{table}

\section{Summary}\label{sec:summary}

We applied a method of the time-dependent AMD to dipole transitions in 
light neutron-rich nuclei. 
We investigated the $E1$ resonances in Be, B, and C isotopes.
It was found that the remarkable peaks appear in $^{10}$Be, $^{15}$B, 
and $^{16}$C at $E_x=10-15$ MeV which almost exhaust the values of the 
cluster sum rule.
These soft dipole resonances arise from the relative motion between
excess neutrons and the core, which is decoupled with the motion inside the
core. In other words, these soft resonances appear due to the
excitation of the excess neutrons around the rather hard core. 
This nature of the neutron excitation and the inert core may have a link with
such ground-band properties of the $^{16}$C as the unusually small 
$B(E2;2^+_1\rightarrow 0^+_1)$\cite{Imai04}, 
which has been recently investigated in theoretical and experimental studies
\cite{ENYO-c16,Elekes04,Sagawa-c16,Takashina}.
In further neutron-rich B and C 
isotopes with $N>10$, the strength for the
soft dipole resonances declines 
compared with that of $^{15}$B and 
$^{16}$C. It is considered to be because the motion 
of the excess neutrons assimilates into the neutron motion within the core.
As a result, the excitation energies of the GDR decrease 
with the enhancement of the neutron-skin.
It is striking that the strength for the soft dipole resonances 
does not necessarily increase with the increase of the excess neutrons.
Instead, the feature of the soft resonances rapidly changes depending on the
proton and neutron numbers of the system. The key of the soft dipole
resonance with a remarkable strength is how the coherent motion 
of the valence neutrons is decoupled with the motion inside the core.

The present method based on the time-dependent AMD 
is regarded as a kind of small-amplitude time-dependent 
Hartree-Fock calculations within the AMD model space.
The point of the method is that we are able to 
study dipole resonances with the framework which can describe 
cluster aspect. In the AMD approach, the dipole excitations are expressed 
by the motion of single-particle Gaussian wave packets,
because the expectation value 
of the dipole operator is related directly to the centers of the 
Gaussian wave packets. 
One of the advantages of the time-dependent
AMD is that we can link the excitations with such collective modes as 
core vibration, core-neutron motion and inter-cluster motion,
which should be important 
to understand the role of the excess neutrons in the dipole resonances.

Recently, extended methods of time-dependent mean-field
calculations have been proposed and applied to the dipole 
transitions in neutron-rich nuclei. 
In the TDHF+ABC approach, which have been applied
to deformed neutron-rich nuclei by Nakatsukasa and Yabana\cite{Nakatsukasa},
the effects of continuum states are taken into account.
Another method is the time-dependent density-matrix theory which 
has been applied to $^{22}$O by incorporating two-body correlations
\cite{Tohyama95}. 
In the present work, the contributions of continuum states 
are omitted, and the detailed descriptions of wave functions and
two-body correlations should be insufficient, 
as the model space is a simple AMD wave function written by
a Slater determinant of Gaussian wave packets. 
Therefore, we put an artificial smoothing parameter
to simulate the width of the dipole resonances, 
because it is difficult to describe the escape and the 
spreading widths of the resonances in the present framework. 
Further extensions of the model
must be essential to give quantitative discussions of the 
excitation energies and the strengths
of the dipole resonances in nuclei near the drip line.
It should be necessary to solve the 
remaining problem of the soft resonances in halo nuclei 
\cite{Honma90,Hoshino91,Sagawa92,Myo98,Suzuki00,Nakamura,Palit}.

We comment that 
the usual AMD wave functions applied to the nuclear structure study
are the advanced ones with some extensions 
such as the parity and spin projections, 
deformed Gaussian base and generator coordinate method
\cite{ENYOsup,Kimura01,ENYOe,Itagaki00,Thiamova03}, though the
present AMD wave function is the simplest one with no extension.
A combination of the time-dependent method and the extended
AMD wave functions should be needed to include higher correlations 
beyond the present calculations, and also to obtain
better description of the ground state properties.
Moreover, interaction dependence of the
dipole transitions is a remaining problem.

\acknowledgments

The authors would like to thank Prof. H. Horiuchi for many discussions.
One of authors, Y. K. is also thankful to Prof. T. Nakatsukasa and Prof. 
M. Tohyama for valuable comments.
Discussions during the workshop YITP-W-05-01 on 
"New Developments in Nuclear Self-Consistent Mean-Field Theories", which was 
held in the Yukawa Institute for 
Theoretical Physics at Kyoto University,
were useful to complete this work. 
The computational calculations in this work were supported by the 
Supercomputer Projects of High Energy Accelerator Research Organization(KEK).
This work was supported by Japan Society for the Promotion of 
Science and a Grant-in-Aid for Scientific Research of the Japan
Ministry of Education, Science and Culture.
A part of the work was performed in the ``Research Project for Study of
Unstable Nuclei from Nuclear Cluster Aspects'' sponsored by
Institute of Physical and Chemical Research (RIKEN).

\end{document}